\documentclass[12pt,peerreview,onecolumn]{IEEEtran}
\usepackage{amsfonts}
\usepackage{amsmath}
\usepackage{centernot}
\usepackage{cite}

\begin{document}

\title{Information Entropy Dynamics and
\\ Maximum Entropy Production Principle}


\author{Alexander~L.~Fradkov,~\IEEEmembership{Fellow,~IEEE,}
Dmitry~S.~Shalymov}



\maketitle

\begin{abstract}
The asymptotic convergence of probability density function (pdf)
and convergence of differential entropy are examined
for the non-stationary processes that follow the maximum entropy principle (MaxEnt) and maximum entropy production principle (MEPP).
Asymptotic convergence of pdf provides
new justification of MEPP
while convergence of differential entropy is important in asymptotic analysis of communication
systems.
A set of equations describing the dynamics of pdf
under mass conservation and energy conservation constraints is derived.
It is shown that for pdfs with compact carrier the limit pdf is unique and can be obtained from Jaynes's MaxEnt principle.
\end{abstract}


\section{Introduction}

Matters of evolution in the nature and society raise a great number of various discussions.
One of the efficient ways to describe and interpret the evolution is to use entropy.

Entropy is an information theory tool, designed to quantify the
information content and possibly information loss for various systems.
The term of information can be used as an inverse of uncertainty.
In this case the growth of uncertainty is identiﬁed with an increase of the entropy which in turn is
interpreted as the information loss.

Entropy and related information measures provide useful descriptions of the long term behavior of random processes.
The notion of entropy often becomes the center of discussions both in statistical physics and in thermodynamics.

To describe the evolution of non-stationary processes that follow the principle of maximum entropy
the maximum entropy production principle (MEPP) is often used \cite{c777}.
The MEPP is that a system evolves in ``natural'' environment maximizing its entropy production when some restrictions are imposed. The ``natural'' environment means that there is no direct external action. Thus if the principle of maximum entropy (MaxEnt) says that a system tends to its state of maximum entropy then the MEPP says that the system tends to it at the highest possible rate. In some cases of system evolution analysis, these principles allow one to obtain missing information required in order to determine the direction of further evolution.

For now, attempts to justify MEPP widely use an approach based on the information entropy proposed by E.T. Jaynes (1957) \cite{c2, c63, c63_2}. This approach is a simple and convenient way to construct (both classical and quantum) statistical thermodynamics; it does not require some complexities such as the ergodic hypothesis. An important advantage of Jaynes' formalism is its ability of being generalized to analyze non-equilibrium systems.

An approach to derive the MEPP using Jaynes' formalism is proposed by Dewar \cite{c77, c78}.
This approach seeks the most probable path of evolution which describes non-equilibrium stationary systems.
Dewar analyzes a non-equilibrium stationary open system (with volume $V$ and boundary $\Omega$) and uses a concept of ``microscopic path'' that implies microstate changes over time. However, according to the discussion in \cite{c777}, it is still an open problem to trace connection between the maximum information entropy of trajectories and the MEPP, and therefore to determine the entropy evolution of the system. We propose a new justification of MEPP based on control theory.


It is known that methods of the optimal control (such as Bellman dynamic programming, Pontryagin's maximum principle, etc.) can be effectively used to develop models of mechanical \cite{c4}, thermodynamic \cite{c5} and other complex systems.
Using such methods also seems promising in the context of the MEPP. Equations of the dynamics of physical systems can be obtained from extremal principles if the goal is to reach the extremum of an objective functional. We can consider the entropy maximization of a system as such a goal and use the speed-gradient (SG) principle \cite{Fradkov_4,Fradkov_5,Fradkov_UFN,Fradkov_7,c10} originated in control theory. In this case, having determined the equations of the system dynamics that ensure the maximum entropy increment, the SG principle can justify the MEPP.

The SG principle is successfully applied in \cite{Fradkov_3, Fradkov_4} to obtain equations of the statistical dynamics of finite systems of particles that follow the principle of maximum entropy. Applicability of the SG principle is experimentally tested in \cite{Fradkov_7}; it is done on the system of particles simulated with the molecular dynamics method on the basis of equations of classical mechanics. However, it is still an open question to apply these results to systems with continuous distributions.

An approach proposed in this paper provides the evolution law of the system in the following form:
\begin{equation}\label{common_eq}
    \dot{p}(t,r) = -\Gamma(I-\Psi)\log{p},
\end{equation}
where $I$ is identity operator, $\Psi$ is a linear integral operator that is invariant for $p,$
$\Gamma>0$ is a constant gain.

By means of this approach the distribution corresponding to
the maximum value of entropy can be found.
A way how to find the distribution achieving
the extreme value (maximum or minimum) of entropy
within a given variational distance from any given distribution is proposed
in \cite{Ho_1}. Approximation of probability distribution is built there based on a new bounds on entropy distance.
Also these bounds are applied to entropy estimation.
The application of these bounds to typical sequences is discussed in \cite{Ho_2}.

Convergence of differential entropies which corresponds to pdfs from eq. (\ref{common_eq})
can be proven on the basis of sufficient conditions provided by Godavarti and Hero \cite{Godavarti}.
According to Godavarti and Hero \cite{Godavarti}, the convergence of differential entropies
can be traced to the problem of asymptotic analysis of various communication systems
where the maximum asymptotic rate of communication is investigated \cite{G_4,G_9,G_16,G_18,G_20}.
Also the differential entropy convergence can be used in communication systems where
the asymptotically optimal source compression \cite{G_7,G_8} or the asymptotic storage capacity \cite{G_14} is analyzed.
Entropy estimation and the convergence of differential entropies are also investigated in \cite{G_3, G_1, G_17}


Jaynes's MaxEnt principle was successfully used
for inductive inference in \cite{Shore}.
It is shown that given information in the form of constraints on expected values,
there is only one appropriate distribution which can be obtained by maximizing entropy.
Appropriateness of this unique distribution is that it satisfies the constraints that can be chosen by a procedure that satisfies the consistency axioms. Continuous probability densities are also considered there.
Inductive inference can be used in systems with non-stationary processes which satisfy the MaxEnt \cite{PhysScripta}.
In \cite{PhysScripta} the notion of Entropy Dynamics (ED) is used.
ED is a theoretical framework which combines inductive inference \cite{Shore}
 and Information Geometry \cite{Amari}. ED investigates the possibility of deriving dynamics from purely entropic arguments.


The main result of this paper is the extension of the \cite{Fradkov_3, Fradkov_4} results to systems with continuous probability distribution. The SG principle is used to derive equations of the dynamics of transient states of the systems that in a steady state follow the principle of maximum entropy.

The next section formulates the SG principle.
Jaynes' formalism is introduced in 3rd section.
The 4th section gives an example of a dynamic system with continuous distribution of parameters that follows the principle of maximum entropy. Cases with one and two constraints are reviewed. Equations of the transient state are derived; and their properties are analyzed. The asymptotic stability of pdfs and differential entropy convergence is proved.

\section{Speed-Gradient Principle}
Let us consider a category of open physical systems which dynamics is described by the system of differential equations
\begin{equation}\label{PrincSkorGrad}
    \dot{x}=f(x,u,t),
\end{equation}

where $x\in\mathbb{C}^n$ is the system state vector, $u$ is the vector of input (free) variables $t\geq 0.$
The task of system simulation (model development) can be defined as deriving the law of variation (of evolution) $u(t)$ that satisfies some criterion of ``naturalness'' of its performance to give the model features characterizing a real physical system.

Developing such a criterion based on extreme and variational principles usually implies specifying some integral functional (for example, an action functional of the principle of least action \cite{Lancosh_8}) that characterizes the system performance. Functional minimization defines probable trajectories of the system $\{x(t),u(t)\}$ as points in the corresponding functional space. The advanced apparatus of the calculus of variations is used to define the law of the system dynamics.

Besides integral principles, differential (local by time) ones are used, for example, Gauss's principle of least constraint, the principle of least dissipation of energy, etc. According to M. Planck \cite{Plank_9}, local principles have some advantage over integral ones because the current state and motion of the system do not depend on its later states and motions there. Using \cite{Fradkov_5, Fradkov_3}, let us formulate one more local variational principle based on the speed-gradient method proposed earlier to synthesize the laws of the nonlinear and adaptive control.

\textbf{The speed-gradient principle:} \emph{of all the probable motions, the system implements the ones for which input variables vary directly as the speed-gradient of some ``goal'' functional} $Q_t.$ If restrictions are imposed on a motion of the system then the direction is a speed-gradient vector projection on the acceptable directions (the ones that satisfy the restrictions) set.

The SG principle can be represented in the finite form
\begin{equation} \label{SkGrad2}
    u=-\Gamma\nabla_{u}\dot{Q_t}.
\end{equation}

where $\dot{Q_t}$ is a rate of change of the goal functional along the trajectory of the system (\ref{PrincSkorGrad}). Let us describe application of the SG principle in the simplest (but at the same time in the most important) case where a category of models of the dynamics (\ref{PrincSkorGrad}) is specified as the relation:
\begin{equation}\label{eq_4}
    \dot{x}=u.
\end{equation}

The relation (\ref{eq_4}) just means that we are deriving the law of variation of variables speeds of the system state. In accordance with the SG principle, the goal functional (function) $Q(x)$ needs to be specified first. $Q(x)$ should be based on physics of a real system and reflect its tendency to decrease the current $Q(x(t))$ value. After that, the law of dynamics can be expressed as (\ref{SkGrad2}).

The SG principle is also applicable to develop models of the dynamics of distributed systems that are described on infinite-dimensional state spaces. There in particular can be a vector $x$ in a Hilbert space $\mathcal{X}$ and a nonlinear operator $f(x,u,t)$ defined on a dense set $D_{F}\subset\mathcal{X}$; in this case, the solutions of equation (\ref{PrincSkorGrad}) are generalized.

\section{Jaynes's Maximum Entropy Principle}
The approach proposed by Jaynes \cite{c2,c63,c63_2} became the foundation for statistical physics nowadays.
Its main ideas are described below.

Let $p(x)$ be a probability density function (pdf) of a multidimensional random variable $x$. This is an unknown function that needs to be defined on the basis of certain system information. Let us suppose that there is the information about some average values $\overline{H}_m$ which are known a priori:
\begin{equation}\label{sred_ogr}
    \overline{H}_m=\int{H_m(x)p(x)dx},~m=1,...,M.
\end{equation}
The next equality is also true for the density function
\begin{equation}\label{plotn_raspred}
    \int{p(x)dx}=1.
\end{equation}

Conditions (\ref{sred_ogr}) and (\ref{plotn_raspred}) in general can be insufficient to derive $p(x)$. In this case, according to Jaynes, applying maximization of information entropy $S_I$ is the most objective method to define the density function.
\begin{equation}\notag
    S_I=-\int{p(x)\log{p(x)}dx}.
\end{equation}

Maximum search with additional conditions (\ref{sred_ogr}) and (\ref{plotn_raspred}) is performed by using Lagrange multipliers; it leads to
\begin{equation}\label{Jaynes_1}
p(x)=\frac{1}{Z}exp\left(-\sum^{M}_{m=1}{\lambda_{m}H_m}\right),
\end{equation}
\begin{equation}\label{Jaynes_2}
Z=\int{exp\left(-\sum^{M}_{m=1}{\lambda_{m}H_m}\right)dx},
\end{equation}
where $\lambda_m$ can be derived from conditions (\ref{sred_ogr}).

These formulas allow definition of the distribution function for microcanonical, canonical and other ensembles.
We substitute the conditions characterizing each equilibrium ensemble for condition (\ref{sred_ogr}) \cite{c63}.
In case of equilibrium (when appropriate random variables x are selected), the formulas also show that the maximum information entropy coincides with the Gibbs entropy and can be identified with the thermodynamic entropy.

Jaynes showed a close connection and succession between his approach and both the classical papers dealing with probability theory and statistics (Bernoulli, Laplace) and the papers doing with physics and information theory (in particular J. Gibbs and C. Shannon) \cite{c65}.

Although information theory was initially based on some concepts of statistical physics, now, thanks to C. Shannon, the information approach can be taken as a basis to develop statistical physics. According to \cite{c777}, formalism of statistical mechanics is some sequence of operations. Having limited information about microworld but following this sequence, we can get the most objective estimation (this is a method of error prevention).

\section{Maximization of entropy with Speed-Gradient method}
The maximum entropy principle of Gibbs and Jaynes \cite{c63,c63_2} defines the asymptotic behavior of the system, but does not say anything about how there is a movement to an asymptotic behavior. To answer this we use the SG principle.
\subsection{System with a continuous distribution of states}
Consider a system with a continuous distribution of the set of possible states.
Probability distribution over states is characterized by pdf $p(t,r)$ which is
continuous everywhere except for a set with zero measure:
\begin{equation}\label{plotnost_ver}
    \int_{\Omega}{p(t,r)dr} = 1,
\end{equation}
where $\Omega$ is a compact carrier.

State of the system evolves over time. We are interested in the behavior of both steady and transient mode. Steady state is determined by the MaxEnt principle. If nothing else about the system is known then its supreme  behavior will maximize its measure of uncertainty (entropy). As a measure of uncertainty we choose differential entropy which is defined as
\begin{equation}\label{diff_entropy}
    S=-\int_{\Omega}{p(t,r)\log{p(t,r})dr}.
\end{equation}

Let's define a law of system dynamics as
\begin{equation}\label{system}
    \dot{x} = u(t,r), x = p(t,r).
\end{equation}
We have to define a function $u(t,r)$.

According to speed-gradient principle a rate of entropy change (\ref{diff_entropy}) which is based on a system (\ref{system}) has to be calculated first. Then a gradient of rate for function $u$ has to be found. And finally, control parameters has to be defined. These parameters are proportional to projection of gradient on a surface of bounds (\ref{plotnost_ver}).

Let's calculate $\dot{S}:$
\begin{equation}
    \dot{S} = -\int_{\Omega}{(u\log{p} + u\frac{p}{p})dr} = -\int_{\Omega}{u\log{p}dr} - \int_{\Omega}{udr}.
\end{equation}

From (\ref{plotnost_ver}) it follows that
\begin{equation}\label{ogr_u}
    \int_{\Omega}{u(t,r)dr} = 0.
\end{equation}
So, $\dot{S} = -\int_{\Omega}{u\log{p}dr}.$
Gradient of $\dot{S}$ by $u$ is equal to $\nabla_{u}{\dot{S}} = -\nabla_{u}{\langle\log{p},u\rangle}$.
According to (\ref{grad_skal_pr}) we have that $\nabla_{u}{\dot{S}} = -\log{p}.$

Speed-gradient principle of motion forms
$u=-\Gamma\log{p(t,r)} + \lambda,$
where $\Gamma$ can be taken as a scalar value and LaGrange multiplier $\lambda$
is selected to satisfy a restriction (\ref{ogr_u}).
\begin{equation}
    \int_{\Omega}{\left(-\Gamma\log{p(t,r)} + \lambda\right)dr}=0 \Rightarrow \lambda = \frac{\Gamma\int_{\Omega}\log{p(t,r)dr}}{\mathrm{mes}(\Omega)},
\end{equation}
where
\begin{equation}\label{mes}
    \mathrm{mes}(\Omega) = \int_{\Omega}{1d\Omega}.
\end{equation}

Final system dynamics equation has the following form
\begin{equation}\label{odno_ogr1}
    \dot{p} = -\Gamma\log{p(t,r)} + \frac{\Gamma\int_{\Omega}\log{p(t,r)dr}}{\mathrm{mes}(\Omega)}=-\Gamma\left(\log{p(t,r)}-\frac{\int_{\Omega}\log{p(t,r)dr}}{\mathrm{mes}(\Omega)}\right).
\end{equation}

Eq. (\ref{odno_ogr1}) can be represented in more general form
\begin{equation}\label{common_form}
    \dot{p} = -\Gamma(I-\Psi)\log{p(t,r)},
\end{equation}
where $\Psi = \frac{\int_{\Omega}{(\cdot)dr}}{\mathrm{mes}(\Omega)}$ is a linear operator which is invariant for $p$
and $I$ ia an identity operator.

Physical meaning of equation (\ref{odno_ogr1}) can be described as a movement to the direction of maximum rate of entropy production. It corresponds to MEPP.

\subsubsection{Equilibrium stability}
Let's investigate a stability of obtained equilibrium equation (\ref{odno_ogr1}).

\textbf{Theorem 1}
\emph{There exists a unique limit pdf $p^{*}(t,r)$ defined by equation} (\ref{odno_ogr1}):

$S(p^{*}(t,r)) = S_{max}.$
\emph{And} $p^{*}(t,r) = \mathrm{mes}^{-1}(\Omega).$

\textbf{Proof}

Consider an entropy function of Lyapunov.
\begin{equation}\label{Lyap_11}
    V(p) = S_{max}-S(p)\geq0.
\end{equation}

Let's find a derivative of function (\ref{w1})
\begin{equation}\label{w1}
    \dot{V}(p)=-\dot{S}(p)=\int_{\Omega}{u\log{p}dr}.
\end{equation}
After substitution of expression for $u$ from (\ref{odno_ogr1}) to (\ref{w1}) we obtain
\begin{equation}\notag
    \dot{V}(p) = \int_{\Omega}{-\Gamma\left(\log^2{p(t,r)}-\frac{\int_{\Omega}{\log{p(t,r)}dr}}{\mathrm{mes}(\Omega)}\log{p(t,r)}\right)dr}.
\end{equation}

After several conversions it is true that
\begin{equation}\notag
    \dot{V}(p) = -\frac{\Gamma}{\mathrm{mes}(\Omega)}\left(\int_{\Omega}{\mathrm{mes}(\Omega)\log^2{p(t,r)}dr}
    -\left(\int_{\Omega}{\log{p(t,r)}dr}\right)^2dr\right).
\end{equation}

We will use the Cauchy-Bunyakovsky inequality
\begin{equation} \label{koshi-bun}
    \left|\int_{\Omega}{f(x)g(x)dx}\right|^2 \leq \left(\int_{\Omega}\left|f(x)\right|^2\right)\left(\int_{\Omega}\left|g(x)\right|^2\right)
\end{equation}
for functions $f=\log{p}$ and $g=1$.
Taking into account that a scalar value $\Gamma$ is positive we get that $\dot{V}(p)\leq0$.
Let's denote a set of functions where $\dot{V}$ has null values as $D=\{p(r,t): \dot{V}=0\}.$
It is known that equality in the Cauchy-Bunyakovsky inequality is achieved when multiplicity occurs, i.e. $f(x)=\alpha g(x).$
In our case $\dot{V}(p)=0$ is true when $\log{p(t,r)} = \alpha.$ It is possible only when $p(t,r) = const.$
Using restriction (\ref{plotnost_ver}) we get that $const = \mathrm{mes}^{-1}(\Omega)$.
And set $D$ consists of only one solution $D=\{p^*\}$.
$\Box$

\subsubsection{Asymptotic convergence}
We will show an asymptotic convergence of all solutions to $p^*.$
To do it we use Barbalat's lemma.

\textbf{Barbalat's lemma}
\emph{If differentiable function $f(t)$ has a finite limit for $t\rightarrow\infty$ and its derivative $\dot{f}(t)$ is uniformly continuous then $\dot{f}(t)\rightarrow0$ for $t\rightarrow0$}

\textbf{Theorem 2}
\emph{For all pdfs defined by equation} (\ref{odno_ogr1})
it is true that $p(t,r)\rightarrow p^*$ for $t\rightarrow \infty.$

\textbf{Proof}

For the sake of simplicity we define a designation for $V$ in (\ref{Lyap_11}) as $v(t)=V\left(p(t)\right).$
We will use a Barbalat's lemma to show that $\dot{v}(t)\rightarrow0.$
We use $v(t)$ as a function $f(t).$
Because of $v(t)\geq0$ and $\dot{v}\leq0$ the function $v(t)$ has a finite limit for $t\rightarrow\infty$.

Let's show conditions when function $\left|\ddot{v}(t)\right|$ is bounded.
That will lead us to the fact that $\dot{v}$ is uniformly continuous.
\begin{multline}
    \ddot{v}(t)
    =-\frac{2\Gamma}{\mathrm{mes}(\Omega)}
    \left(\mathrm{mes}(\Omega)-2\int_{\Omega}{\log^2{p}dr}\right)\int_{\Omega}{\log{p}\frac{\dot{p}}{p}dr} =
    \\
    = \frac{2\Gamma^2}{\mathrm{mes}(\Omega)}\left(\mathrm{mes}
    (\Omega)-2\int_{\Omega}{\log^2{p}dr}\right)\left[\int_{\Omega}{\frac{\log^2{p}}{p}dr}-\int_{\Omega}{\frac{\log{p}}{p}dr}\frac{\int_{\Omega}{log{p}dr}}{\mathrm{mes}(\Omega)}\right].
\end{multline}

Expression $\int_{\Omega}{\log^2{p}dr}<\infty$ comes true when
\begin{equation}\label{mes_0}
    \mathrm{mes}\{{r:p(r)=0\cup p(r)=\infty}\} = 0,
\end{equation}
where $\mathrm{mes}$ is defined as (\ref{mes}).

A similar conclusion may by performed for function $\ddot{v}(t).$
In this way a function $\dot{v}$ is uniformly continuous if condition (\ref{mes_0}) is satisfied.
Then according to Barbalat's lemma $\dot{v}(t)\rightarrow0$ for $t\rightarrow0$.

Let $f_t=\log{p_t}$. Taking into account that $\int_{\Omega}{f^2dr}=\|f\|^2$ the expression for $\dot{v}$ may be written as
\begin{multline} \label{dot_v}
    \dot{v}=-\frac{\Gamma}{\mathrm{mes}(\Omega)}\left[\mathrm{mes}(\Omega)\|f_t\|^2-(1,f_t)^2\right]
    =-\Gamma\|f_t\|^2\left[1-\frac{(1,f_t)^2}{\mathrm{mes}(\Omega)\|f_t\|^2}\right] = \\
    =-\Gamma\|f_t\|^2(1-\cos^2(\alpha(t))),
\end{multline}
where $\alpha(t)=\cos(\langle 1, f_t\rangle).$

If $\|f_t\|^2\rightarrow0$ then $\int_{\Omega}{\log^2{p_t}dr}\rightarrow0$.
It means that $\mathrm{mes}_t\{r: p_t(r)\neq 1\}\rightarrow0.$ It is equivalent to $\int_{\Omega}{p_tdr}\rightarrow \mathrm{mes}(\Omega).$
Due to the bound for the density for all $\forall t$ it is true that $\int_{\Omega}{p_tdr} = 1.$
Thus $\|f_t\|^2\rightarrow0$ is wrong if only $\mathrm{mes}(\Omega)\neq 1.$
The case when $\mathrm{mes}(\Omega)=1$ we consider as a degenerate.
Given (\ref{dot_v}) we obtain $\alpha(t)\rightarrow0.$
It means that $\widehat{f_t}\rightarrow\widehat{1},$ where $\widehat{f_t}$ and $\widehat{1}$ are
normalized values for $f_t$ and $1$ respectively.

It follows that $p_t$ tends to the stationary distribution. As explained earlier this distribution is unique.

Thus, $p_{t_{t\rightarrow\infty}}\rightarrow p^*.$
$\Box$

\subsubsection{Convergence of differential entropies}
Let's prove the convergence of differential entropies

\textbf{Theorem 3}
\emph{For differential entropy defined as
\begin{equation}\notag
H(t) = H(p(t,r))=-\int_{\Omega}{p(t,r)\log{p(t,r)dr}}
\end{equation}
it is true that
\begin{equation}\label{entr_conv}
H(t)_{t\rightarrow\infty}\rightarrow H(p^*),
\end{equation}
where pdf $p(t,r)$ is defined by equation (\ref{odno_ogr1})
and $p^{*}$ is limiting pdf:
$p(t,r)_{t\rightarrow\infty}\rightarrow p^{*}(r).$
}

\textbf{Proof}

In \cite{Godavarti} Godavarti and Hero provide sufficient conditions for convergence of differential entropies $H(p_n)\rightarrow H(p_*)$.
The only assumptions are that $p_n\rightarrow p_*$
and that exists a constant $L$ and some $k>1$ such that for all $N$
\begin{equation}\label{g1}
    \int{p_{N}|\log{p_{N}}|^k}\leq L
\end{equation}
and
\begin{equation}\label{g2}
    \int{p|\log{p}|^k}\leq L
\end{equation}

For all $t$ our pdf $p(t,r)$ from (\ref{odno_ogr1})
the first assumption $p_n\rightarrow p_*$ is satisfied due to Theorem 2.
Conditions (\ref{g1}) and (\ref{g2}) are also satisfied
on the compact carrier $\Omega.$
According to \cite{Godavarti} it means that (\ref{entr_conv}) is true.
$\Box$

So with eq. (\ref{common_form})
we are able to predict the evolution of system's entropy.

\subsection{Total energy constraint}
Restriction (\ref{plotnost_ver}) can be interpreted as the law of conservation of mass on a space $\Omega.$
Let's consider a system with additional constraint for the law of total energy conservation.
We will consider a conservative case when energy does not depend on a time.
The new constraint may be described as
\begin{equation}\label{ogr_energ}
    \int_{\Omega}{p(t,r)}h(r)dr=E,
\end{equation}
where $E$ is a common energy of a system and $h(r)$ is a density of energy.

Let's consider a system
\begin{equation}\label{system_eq}
    \dot{p} = u.
\end{equation}

The problem is to find an operator $u$ which satisfies to both constraints (\ref{plotnost_ver}) and (\ref{ogr_energ}) at any time $t$.
And the following goal condition is true.
\begin{equation}\notag
    S\left(p(t,r)\right)_{t\rightarrow\infty}\rightarrow S_{max}.
\end{equation}

To solve this problem we use the speed-gradient method. As a goal function we take
\begin{equation}\notag
Q(p)=S_{max}-S+\lambda'_1\left(\int_{\Omega}{p(t,r)h(r)dr}-E\right)+\lambda'_2\left(\int_{\Omega}{p(t,r)dr}-1\right),
\end{equation}
where $S$ is a differential entropy, $\lambda'_1$ and $\lambda'_2$ are LaGrange multipliers.
Due to the speed-gradient principle an operator $u$ has to be described as
\begin{equation}\notag
    u=-\Gamma\nabla_u{\dot{Q}(p,u,t)}.
\end{equation}

Let's calculate the derivative of a goal functional
\begin{multline}\notag
    \dot{Q}=\int_{\Omega}{\dot{p}(t,r)\log{p(t,r)}dr} + \int_{\Omega}{\dot{p}(t,r)dr} + \lambda'_1\left(\int_{\Omega}{\dot{p}(t,r)h(r)dr}\right) + \lambda'_2\int_{\Omega}{\dot{p}(t,r)dr} = \\
    \int_{\Omega}{u(t,r)\log{p(t,r)}dr} + \lambda'_1\int_{\Omega}{u(t,r)h(r)dr} + \left(\lambda'_2+1\right)\int_{\Omega}{u(t,r)dr}.
\end{multline}

Taking the gradient by $u$ we get
\begin{equation}\notag
    \nabla_u{\dot{Q}} = \log{p(t,r)} + \lambda'_{1}h(r) + \left(\lambda'_1+1\right)u(t,r).
\end{equation}

Now then
\begin{equation}\label{dva_ogr}
    u = -\Gamma\log{p(t,r)} + \lambda_{1}h(r) + \lambda_2,
\end{equation}
where $\lambda_1=-\Gamma\lambda'_1, $ $\lambda_2=-\Gamma(\lambda'_2+1)$.

Now we find LaGrange multipliers $\lambda_1$ and $\lambda_2$ based on boundary conditions.
For the sake of simplicity we omit the arguments of functions. So, for example we use $\log{p}$ instead of $\log{p(t,r)}.$
It is follows from the condition (\ref{ogr_u}) that
\begin{equation}\label{1usl}
    -\Gamma\int_{\Omega}{\log{p}dr}+\lambda_1\int_{\Omega}{{h}dr}+\lambda_{2}\mathrm{mes}(\Omega) = 0.
\end{equation}

Condition (\ref{ogr_energ}) is equivalent to $\int_{\Omega}{\mathrm{uh}dr}.$ From this condition it should be
\begin{equation}\label{2usl}
    -\Gamma\int_{\Omega}{\log{(p)}hdr} + \lambda_1\int_{\Omega}{h^2dr} + \lambda_2\int_{\Omega}{hdr}.
\end{equation}
Solving the system of equations (\ref{1usl}) and (\ref{2usl}) we obtain
\begin{equation}\label{koeff1}
    \lambda_1=\Gamma\frac{\mathrm{mes}(\Omega)\int_{\Omega}{\log{(p)}hdr}-\int_{\Omega}{\log{(p)}dr}\int_{\Omega}{hdr}}{\mathrm{mes}(\Omega)\int_{\Omega}{h^2dr}-\left(\int_{\Omega}{hdr}\right)^2},
\end{equation}
\begin{equation}\label{koeff2}
   \lambda_2=\Gamma\frac{\int_{\Omega}{\log{(p)}dr}\int_{\Omega}{h^2dr}-\int_{\Omega}{hdr}\int_{\Omega}{\log{(p)}hdr}}
    {\mathrm{mes}(\Omega)\int_{\Omega}{h^2dr}-\left(\int_{\Omega}{hdr}\right)^2}.
\end{equation}

Equations (\ref{koeff1}) and (\ref{koeff2}) are defined when denominator in both fractions is not equal to zero.
If we take in the Cauchy-Bunyakovsky inequality $f=h$ аnd $g=1$ then the following inequality comes true
\begin{equation}\label{koshi_bun_int}
    \left|\int_{\Omega}{hdr}\right|^2 \leq \mathrm{mes}(\Omega)\int_{\Omega}{h^2dr},
\end{equation}
This inequality becomes equality when $h=const.$ It means that all energy levels coincide.
This case is supposed to be degenerate and not considered here.
Thus the expression
\begin{equation}
\mathrm{mes}(\Omega)\int_{\Omega}{h^2dr}\neq\left(\int_{\Omega}{hdr}\right)^2.
\end{equation}
is always true.

Given (\ref{koeff1}), (\ref{koeff2}) and substituting (\ref{dva_ogr}) in equation (\ref{system_eq})
we obtain the common equation of system dynamics:
\begin{multline}\label{dva_ogr_full}
    \dot{p}(t,r) = -\Gamma\log{p(t,r)} + \Gamma\frac{\mathrm{mes}(\Omega)\int_{\Omega}{\log{(p)}hdr}-\int_{\Omega}{\log{(p)}dr}\int_{\Omega}{hdr}}{\mathrm{mes}(\Omega)\int_{\Omega}{h^2dr}-\left(\int_{\Omega}{hdr}\right)^2}
h(r) + \\
\Gamma\frac{\int_{\Omega}{\log{(p)}dr}\int_{\Omega}{h^2dr}-\int_{\Omega}{hdr}\int_{\Omega}{\log{(p)}hdr}}
    {\mathrm{mes}(\Omega)\int_{\Omega}{h^2dr}-\left(\int_{\Omega}{hdr}\right)^2}.
\end{multline}

Law of evolution (\ref{dva_ogr_full}) can be represented in abbreviated form:
\begin{equation}\label{common_form}
    \dot{p}=-\Gamma(I-\Psi)\log{p},
\end{equation}
where $I$ is an identity operator,
$\Psi$ is a linear integral operator that is independent of $p$
\begin{equation}\label{IntOperator}
    \Psi = \frac{(1,\cdot)}{\mathrm{mes}(\Omega)} + \frac{\tilde{h}(\tilde{h},\cdot)}{\|h\|^2-\frac{1}{\mathrm{mes}(\Omega)}\left(1,h\right)^2},
\end{equation}
$\tilde{h}=h-\frac{1}{\mathrm{mes}(\Omega)}\int_{\Omega}{hdr}$.

\subsubsection{Stability of equilibrium}
Further we examine the equilibrium of obtained equation (\ref{dva_ogr_full}).

\textbf{Theorem 4}
\emph{There exists a unique limit pdf $p^{*}$ defined by equation} (\ref{dva_ogr_full}):~
$S(p^{*}) = S_{max}.$

\emph{$p^{*} = Ce^{-\mu h},$
where $\mu=-\frac{\lambda_1}{\Gamma},$ $C=e^{\frac{\lambda_2}{\Gamma}}$
and $\lambda_1$, $\lambda_2$ are defined in} (\ref{dva_ogr}).

\textbf{Proof}

Let's calculate the derivative of functional (\ref{Lyap_11}) for the system (\ref{dva_ogr_full}):

\begin{math}
    \dot{V} = \int_{\Omega}{\dot{p}\log{p}dr} = \int_{\Omega}{\left(-\Gamma\log{p}+\lambda_{1}h+\lambda_{2}\right)}\log{p}dr =\\
    = -\Gamma\int_{\Omega}{\log^2{p}dr} + \lambda_1\int_{\Omega}{h\log{p}dr} + \lambda_2\int_{\Omega}{\log{p}dr} = \\
    = -\Gamma\int_{\Omega}{\log^2{p}dr} + \lambda_1\int_{\Omega}{h\log{p}dr} + \frac{\Gamma\int_{\Omega}{\log{p}dr}-\lambda_1\int_{\Omega}{hdr}}{\mathrm{mes}(\Omega)}\int_{\Omega}{\log{p}dr} = \\
    = \frac{\Gamma}{\mathrm{mes}(\Omega)}\left[
    \left( \left( \int_{\Omega}{\log{p}dr}\right)^2 - \mathrm{mes}(\Omega)\int_{\Omega}{\log^2{p}dr} \right) +
    \frac{\left(\mathrm{mes}(\Omega)\int_{\Omega}{h\log{p}dr}-\int_{\Omega}{hdr}\int_{\Omega}{\log^2{p}dr}\right)^2}{\mathrm{mes}(\Omega)\int_{\Omega}{h^2dr} - \left(\int_{\Omega}{hdr}\right)^2}
    \right].
\end{math}

It can be proved that
\begin{equation}\label{neravenstvo}
    \dot{V}(p)\leq 0.
\end{equation}

For the proof of inequality (\ref{neravenstvo}) we define a functional:

$\langle\cdot,\cdot\rangle:L_{2}(\Omega)\times L_{2}(\Omega)\rightarrow\mathbb{R},$ $\forall f,g\in L_{2}(\Omega)$
\begin{equation}
 \langle f,g \rangle = \mathrm{mes}(\Omega)\int_{\Omega}{fgdr} - \int_{\Omega}{fdr}\int_{\Omega}{gdr}.
\end{equation}

New functional has several useful properties

1. Linearity in the first argument

$
\forall f,g,h \in L_{2}(\Omega),~ \forall \lambda\in\mathbb{R}~
\langle\lambda f + g,h\rangle = \langle\lambda f,h\rangle + \langle g,h\rangle.
$

2. Symmetry

$
\forall f,g \in L_{2}(\Omega)~
\langle f,g\rangle = \langle g,f\rangle.
$

3. Positiveness and the condition of zero value

$
\forall f \in L_{2}(\Omega)~
\langle f,f\rangle \geq 0,~ \langle f,f\rangle = 0 \Leftrightarrow f=\mu=const.
$

Let's prove inequality (\ref{neravenstvo}) base on properties 1-3.


Obvious that for any $f,g\in L_{2}(\Omega)$ and $\lambda\in\mathbb{R}$ it is true that $f-\lambda g\in L_{2}(\Omega)$.
This function has property 3: $\langle f-\lambda g, f-\lambda g\rangle\geq 0$.
Using properties 1 and 2 we get the quadratic inequality with respect to $\lambda$:
\begin{equation}\notag
    0\leq\langle f-\lambda g,f-\lambda g\rangle=\lambda^2\langle g,g\rangle-2\lambda\langle f,g\rangle + \langle f,f\rangle.
\end{equation}

This inequality holds for any real $\lambda$.
Hence the discriminant can not be positive.
\begin{equation}\notag
    D = 4\langle f,g\rangle^2 - 4\langle f,f\rangle\langle g,g\rangle.
\end{equation}
Thus
\begin{equation}\label{ner_inner}
    \langle f,g\rangle^2 \leq \langle f,f\rangle\langle g,g\rangle.
\end{equation}

If the equality takes place in (\ref{ner_inner}) then there exists a unique solution $\lambda\in\mathbb{R}$ of an equation
$\langle f-\lambda g, \langle f-\lambda g\rangle = 0$.
But then by property 3 we have
$\exists\mu\in\mathbb{R}:~f-\lambda g=\mu\mathbf{1}$.

Substituting $f=\log{p}, g=h$ to the inequality (\ref{ner_inner}) we get
\begin{multline}\notag
    \left(\mathrm{mes}(\Omega)\int_{\Omega}{h\log{p}dr}-\int_{\Omega}{hdr}\int_{\Omega}{\log{p}dr}\right)^2\leq \\
    \left(\mathrm{mes}(\Omega)\int_{\Omega}{\log^2{p}dr} - \left(\int_{\Omega}{\log{p}dr}\right)^2\right)
    \left(\mathrm{mes}(\Omega)\int_{\Omega}{h^2{p}dr} - \left(\int_{\Omega}{hdr}\right)^2\right).
\end{multline}
Which implies the inequality (\ref{neravenstvo}).

Note that the equality (\ref{neravenstvo}) holds if and only if
\begin{equation}\label{ravenstvo_skal_pr}
    \exists\lambda,\mu\in\mathbb{R}:~\log{p}=\lambda h + \mu.
\end{equation}

In the case of equilibrium the expression (\ref{dva_ogr}) can be written as
\begin{equation}\notag
    \log{p(t,r)}=\frac{\lambda_1 h(r)}{\Gamma} + \frac{\lambda_2}{\Gamma},
\end{equation}
which coincides with (\ref{ravenstvo_skal_pr}).
Thus there is a unique pdf corresponding to the equilibrium state:
\begin{equation}
    p^*=Ce^{-\mu h},
\end{equation}
where $\mu=-\frac{\lambda_1}{\Gamma},$ $C=e^{\frac{\lambda_2}{\Gamma}}$.
This state corresponds to the Gibbs distribution, which is consistent with the results of classical thermodynamics.
$\Box$

\subsubsection{Asymptotic convergence}
We prove asymptotic convergence with Theorem 5.

\textbf{Theorem 5}
\emph{For all pdfs defined by equation} (\ref{dva_ogr_full}) \emph{it is true that}
$p(t,r)\rightarrow p^*$ for $t\rightarrow \infty.$

\textbf{Proof}

The proof of asymptotic convergence for eq. (\ref{dva_ogr_full}) will be performed similar to the case with one constraint (Theorem 2).
To use Barbalat's lemma we have to check conditions under which the function $\ddot{v}$ is limited.
\begin{multline}\label{v_dot}
     \dot{v} = \frac{\Gamma}{\mathrm{mes}(\Omega)}
    \left( \left( \int_{\Omega}{\log{p}dr} \right)^2 - \mathrm{mes}(\Omega)\int_{\Omega}{\log^2{p}dr} \right) + \\ +
    \frac{\Gamma}{\mathrm{mes}(\Omega)}
    \frac{\left(\mathrm{mes}(\Omega)\int_{\Omega}{h\log{p}dr}-\int_{\Omega}{hdr}\int_{\Omega}{\log{p}dr}\right)^2}{\mathrm{mes}(\Omega)\int_{\Omega}{h^2dr} - \left(\int_{\Omega}{hdr}\right)^2}.
\end{multline}

We obtain
\begin{multline}\label{v_ddot}
\ddot{v} =  \frac{2\Gamma}{V}
\left(
\left(\int_{\Omega}{\log{p}dr}\int_{\Omega}{\frac{\dot{p}}{p}dr}-\mathrm{mes}(\Omega)\int_{\Omega}{\log{p}\frac{\dot{p}}{p}dr}\right)
\right) + \\ \frac{2\Gamma}{V}\left(
\frac{\left(\mathrm{mes}(\Omega)\int_{\Omega}{h\frac{\dot{p}}{p}dr}-\int_{\Omega}{hdr}\int_{\Omega}{\frac{\dot{p}}{p}dr}\right)
\left(\mathrm{mes}(\Omega)\int_{\Omega}{h\log{p}dr}-\int_{\Omega}{hdr}\int_{\Omega}{\log{p}dr}\right)}
{V\int_{\Omega}{h^2dr} - \left(\int_{\Omega}{hdr}\right)^2}
\right).
\end{multline}

Considering (\ref{dva_ogr_full}) we get that $\ddot{v}$ is limited under the same conditions which were used in the case of one restriction.

Namely, $\mathrm{mes}\{{r:p(r)=0\cup p(r)=\infty}\} = 0.$

According to Barbalat's lemma it is true that
\begin{equation}\label{v_dot_null}
    \dot{v}\rightarrow0.
\end{equation}
We introduce a scalar product as
$\langle f,f\rangle = \mathrm{mes}(\Omega)\int_{\Omega}{f^2dr} - (\int_{\Omega}{fdr})^2$.
Having $\|f\|^2=\langle f,f\rangle$ the expression (\ref{v_dot}) can be written as
\begin{multline}\label{dva_ogr_skal_pr}
    \dot{v}=\frac{\Gamma}{\mathrm{mes}(\Omega)}\left[-\|\log{p}\|^2+\frac{(\log{p},h)^2}{\|h\|^2}\right]=
    -\frac{\Gamma}{\mathrm{mes}(\Omega)}\|\log{p}\|^2\left[1-\frac{(\log{p},h)^2}{\|\log{p}\|^2\|h\|^2}\right].
\end{multline}

Consider the case when $\|\log{p}\|^2\rightarrow0$.
By Cauchy-Bunyakovsky inequality the equality in inequality
\begin{equation}\notag
    \left(\int_{\Omega}{\log{p}dr}\right)^2\leq \mathrm{mes}(\Omega)\int_{\Omega}{\log^2{p}dr}
\end{equation}
takes place when $\log{p}=\alpha$.

We have previously demonstrated that the equality $\dot{v}=0$ holds in the only one case when $\log{p}=\lambda h + \mu,$ where $\lambda$ and $\mu$ are constants.
Then $h$ must be a constant since the equality $\lambda h + \mu = \alpha.$
Such a case we assume degenerate because then the denominator in the expressions for the coefficients $\lambda_1$ and  $\lambda_2$ ((\ref{koeff1}) и (\ref{koeff2})) is equal to zero.

Given (\ref{v_dot_null}), (\ref{dva_ogr_skal_pr}) and $\|\log{p}\|^2\centernot\rightarrow0$ we obtain $\frac{(\log{p},h)^2}{\|\log{p}\|^2\|h\|^2}\rightarrow 1$.
Which implies that
$\widehat{\log}{p}\rightarrow\widehat{h},$
where $\widehat{\log}{p}$ and $\widehat{h}$ are normalized values for $\log{p}$ and $h$ respectively.
Thus $p$ tends to the only one stationary distribution $p^*$ since $h$ does not depend on time.
$\Box$

\subsubsection{Convergence of differential entropies}
Similar to differential entropies convergence described in paragraph 4.1.3 (see Theorem 3)
the convergence of differential entropies for the case of two constraints can be proved.

\textbf{Theorem 6}
\emph{For pdf which is defined by equation} (\ref{dva_ogr_full})
\emph{it is true that}
\begin{equation}\notag
H(t)_{t\rightarrow\infty}\rightarrow H(p^*),
\end{equation}
\emph{where $H(t)=H(p(t,r))$ is differential entropy and $p^{*}$ is limiting pdf:}
$p(t,r)_{t\rightarrow \infty}\rightarrow p^{*}(r).$

\section{Conclusion}
The MEPP is widely used in different studies of complex systems of physical, chemical or biological origin \cite{c777}.
Using MEPP, P. Zupanovic \cite{c156} derived Kirchhoff law of electrical circuits.

The task to justify MEPP was unresolved for a long time.
We propose such a justification based on results originated in the control theory.

It is an important problem to derive the principle that determines the dynamics of non-stationary (transient) states, and describes a way and trajectory of the system that tends to the state with maximum entropy.
If a goal function is the entropy of a system then the extreme SG principle supplements Gibbs and Jaynes maximum entropy principle to determine the direction of evolution of the system when it tends to the state of maximum entropy. This direction corresponds to maximum entropy production rate (growth rate).

SG principle generates equations for the transient (non-stationary) states of the system operation, i.e. it gives an answer to the question of \textbf{How the system will evolve?} This fact distinguishes the SG principle from the principle of maximum entropy, the principle of maximum Fisher information and others characterizing the steady-state processes and providing an answer to the questions of \textbf{To where?} and \textbf{How far?}

The equations derived on the basis of the SG principle allow forecasting the dynamics of non-equilibrium systems with continuous distribution of parameters. These equations can prove to be useful to study both evolutions and relaxations of non-equilibrium systems of macroscopic and microscopic world.

\appendix
{\bf Preliminary Information}

\emph{Scalar Product.}
In space of real functions $L_2(\Omega)$ a scalar product can be specified as
\begin{equation}\label{skal_pr}
    \langle f,g \rangle = \int_{\Omega}{f(x)g(x)d\Omega},
\end{equation}
where $\Omega$ is a compact carrier. This scalar product generates a norm:
\begin{equation}\label{norma}
    \|f\|_2=\sqrt{\langle f, f \rangle}.
\end{equation}
Allowing completeness of space $L_p(\Omega)$ we can consider space $L_2(\Omega)$ as a Hilbert space.

\emph{Variational Derivative and Gradient of Scalar Product.}
Let us assume that functional $\Phi$ is specified as $\Phi:L_2(\Omega)\rightarrow\mathbb{R}.$ A variational derivative of $\Phi$ can be specified as the analog of a finite-dimensional gradient.
\begin{equation}\notag 
    d\Phi = (\frac{d\Phi}{df},df) = \int_{\Omega}{\frac{d\Phi}{df}(x)df(x)dx},
\end{equation}
where $\frac{d\Phi}{df}$ is a variational derivative of $\Phi$ with respect to $f$, and $df$ is a variation. If variation $d\Phi = \Phi(f + df) - \Phi(f)$ can be expressed as $d\Phi = (A,df) = \int_{\Omega}A(x)df(x)dx$ then $A$ is a variational derivative of $\Phi$ with respect to $f$.

Thus for each scalar product $(a,g) = \int_{\Omega}a(x)g(x)dx$
in space of real functions $L_2(\Omega)$ a function $a$ can be regarded as a variational derivative with respect to $g$, i.e.
\begin{equation}\label{grad_skal_pr}
    \nabla_{g}(a,g) = a.
\end{equation}

\emph{Differential and information entropy.}
If discrete random variable $X$ is $x_1,x_2,...,x_n$~ with with probabilities $p_1,p_2,...,p_n,$~ respectively then its measure of uncertainty is
\begin{equation}\label{1}
  H(X)=-\sum_{i=1}^n {p_i\log(p_i)}.
  \end{equation}

Let us consider $p(x)dx$ as a probability of random variable hitting into indefinitely small interval ~$(x, x+dx)$ then the entropy in \eqref{1} can be formalized for a continuous case as:
\begin{multline}\label{3}
    H(X)=-\int_{-\infty}^{\infty}{\left(p(x)dx\right)\log\left(p(x)dx\right)} = -\int_{-\infty}^{\infty}{p(x)\log\left(p(x)\right)dx} - \int_{-\infty}^{\infty}{p(x)\log\left(dx\right)dx},
\end{multline}
where $p(x)$ is a pdf.

In expression \eqref{3} only the first term depends on the probability density of a random variable but the second one is always ~$\infty$.
The measures of uncertainty for different distributions can be compared by comparing only the first terms of expression \eqref{3}. To do it, a concept of differential measure of uncertainty (differential entropy) is introduced.
\begin{equation}\notag 
    h(X)=-\int_{-\infty}^{\infty}p(x)\log\left(p(x)\right)dx.
\end{equation}

Differential entropy as an entropy of continuous distribution can be interpreted as the average information of continuous source, i.e. information entropy.


\textbf{Properties of functional $\langle f,g\rangle$ from the proof of Theorem 2 (sect. 4.2.2)}

1. Linearity in the first argument

$
\forall f,g,h \in L_{2}(\Omega),~ \forall \lambda\in\mathbb{R}~
\langle\lambda f + g,h\rangle = \langle\lambda f,h\rangle + \langle g,h\rangle.
$

\textbf{Proof}
Using the linearity of the integral we obtain
\begin{multline}\notag
    \langle\lambda f + g,h\rangle = \mathrm{mes}(\Omega)\int_{\Omega}{(\lambda f+g)hdr} - \int_{\Omega}{(\lambda f + g)dr}\int_{\Omega}{hdr} = \\
    = \left(\mathrm{mes}(\Omega)\int_{\Omega}{\lambda fhdr} - \int_{\Omega}{\lambda fdr}\int_{\Omega}{hdr}\right) +
    \left( \mathrm{mes}(\Omega)\int_{\Omega}{ghdr}-\int_{\Omega}{gdr}\int_{\Omega}{hdr}\right) = \\
    = \langle\lambda f,h\rangle + \langle\lambda g,h\rangle.
    ~\Box~~~~~~~~~~~~~~~~~~~~~~~~~~~~~~~~~~~~~~~~~~~~~~~~~~~~~~~~~~~~~~~~~~~~~~~
\end{multline}

2. Symmetry

$
\forall f,g \in L_{2}(\Omega)~
\langle f,g\rangle = \langle g,f\rangle.
$

\textbf{Proof}
\begin{equation}\notag
    \langle f,g\rangle = \mathrm{mes}(\Omega)\int_{\Omega}{fgdr}-\int_{\Omega}{fdr}\int_{\Omega}{gdr} =
    \mathrm{mes}(\Omega)\int_{\Omega}{gfdr}-\int_{\Omega}{gdr}\int_{\Omega}{fdr} = \langle g,f\rangle.~\Box
\end{equation}

3. Positiveness and the condition of zero value

$
\forall f \in L_{2}(\Omega)~
\langle f,f\rangle \geq 0,~ \langle f,f\rangle = 0 \Leftrightarrow f=\mu=const.
$

\textbf{Proof}
Let's consider a scalar multiplication (\ref{skal_pr}).
Cauchy-Bunyakovsky comes true for it:
$\left|(f,g)\right|^2\leq(f,f)(g,g)$ и $\left|(f,g)\right|^2=(f,f)(g,g) \Leftrightarrow \exists\mu\in\mathbb{R}:f=\mu g.$
Substituting $g\equiv1$ we get
\begin{equation}\notag
    \left|\int_{\Omega}{fdr}\right|^2\leq(f,f)\int_{\Omega}{1dr}=\mathrm{mes}(\Omega)\int_{\Omega}{f^2dr}.
\end{equation}
Thereby it is true that
\begin{equation}\notag
    \langle f,f\rangle = \mathrm{mes}(\Omega)\int_{\Omega}{f^2dr} - \left(\int_{\Omega}{fdr}\right)^2 \geq 0.
\end{equation}
Moreover $\langle f,f\rangle=0\Leftrightarrow\left|\int_{\Omega}{fdr}\right|^2=(f,f)(\mathbf{1},\mathbf{1})\Leftrightarrow
\exists\mu\in\mathbb{R}:f=\mu\mathbf{1},$ т.е. $f=\mu$~~~$.\Box$



\begin{thebibliography}{50}

\bibitem{c777}
L.M.Martyushev, V.D. Seleznev,
``Maximum entropy production principle in physics, chemistry and biology''
Phys. Report, V.426 (1), pp. 1--45, 2006.

\bibitem{c2}
Jaynes E.T., ``The minimum entropy production principle''
Ann. Rev. Phys. Chem, V. 31, pp. 579--601, 1980.

\bibitem{c63}
Jaynes E.T., ``Information theory and statistical mechanics''
 Phys. Rev, V. 106. pp. 620--630, 1957;

\bibitem{c63_2}
Jaynes E.T., ``Information theory and statistical mechanics.2''
Phys. Rev, V. 108. pp. 171--190, 1957.

\bibitem{c77}
Dewar R., ``Information theory explanation of the fluctuation theorem, maximum entropy production and self-organized criticality in non-equilibrium stationary states''
J. Phys. A: Math. Gen, V. 36. pp. 631--641, 2003.

\bibitem{c78}
Dewar R., ``Maximum entropy production and the fluctuation theorem''
J. Phys. A: Math. Gen, V. 38, pp. 371--381, 2005.

\bibitem{c4}
Bellman R., Dreyfus S., ``Applied Dynamic Programming Princeton'' New Jersey, 1962.

\bibitem{c5}
Tsirlin A.M., Mironova V.A., Amelkin S.A., Kazakov V.,
``Finite-time thermodynamics: Conditions of minimal dissipation for thermodynamic process with given rate''
Phys. Rev. E, V.58, pp. 215--223, 1998.

\bibitem{Fradkov_4}
Fradkov A L., ``Speed-gradient entropy principle for nonstationary processes'' Entropy, Vol. 10(4), pp. 757-764, 2008.

\bibitem{Fradkov_5}
Fradkov A.L., Miroshnik I.V., Nikiforov, V.O.,
``Nonlinear and Adaptive Control of Complex Systems''
Kluwer Acad. Publ., 1999.

\bibitem{Fradkov_UFN}
Fradkov A. L., ``Application of cybernetic methods in physics'' Physics-Uspekhi, V.48, pp. 103--127, 2005.

\bibitem{Fradkov_7}
Fradkov A., Krivtsov A.,
``Speed-gradient principle for description of transient dynamics in systems obeying maximum entropy principle''
Bayesian inference and maximum entropy methods in science and engineering,
 V. 1305, pp. 399--406, 2010.

\bibitem{c10}
Fradkov A. L., ``Speed-Gradient scheme and its application in adaptive-control problems''
Automation and Remote Control, Vol. 40(9), pp. 1333--1342, 1979.

\bibitem{Fradkov_3}
Fradkov A. L.,
``Cybernetical physics: from control of chaos to quantum control'' Springer-Verlag, 2007.

\bibitem{Ho_1}
S.-W. Ho and R. W. Yeung, ``The Interplay Between Entropy and
Variational Distance'' IEEE Trans. Inform. Theory, vol. 56, no. 12, pp. 5906--5929, Dec. 2010.

\bibitem{Ho_2}
S.-W. Ho and R. W. Yeung, ``On information divergence measures and
a unified typicality'' IEEE Trans. Inform. Theory, vol. 56, no. 12, pp. 5893--5905, Dec. 2010.

\bibitem{Godavarti}
M. Godavarti, A. Hero, ``Convergence of differential entropies''
IEEE Trans. Inform. Theory, vol. 50, no. 1, pp. 171--176, Jan. 2004

\bibitem{G_4}
C.-N. Chuah, D.N. C. Tse, J. M. Kahn, and R. A. Valenzuela, ``Capacity
scaling in mimo wireless systems under correlated fading'' IEEE Trans.
Inform. Theory, vol. 48, pp. 637--650, Mar. 2002

\bibitem{G_9}
A. Lozano and A. M. Tulino, ``Capacity of multiple-transmit multiplereceive
antenna architectures,'' IEEE Trans. Inform. Theory, vol. 48, pp.
3117--3128, Dec. 2002.

\bibitem{G_16}
X. Mestre, J. R. Fonollosa, and A. Pages-Zamora, ``Capacity of MIMO
channels: Asymptotic evaluation under correlated fading'' IEEE J. Select.
Areas Commun., vol. 21, pp. 829--838, June 2003.

\bibitem{G_18}
V. V. Prelov and E. C. van der Meulen, ``An asymptotic expression for
the information and capacity of a multidimensional channel with weak
input signals'' IEEE Trans. Inform. Theory, vol. 39, pp. 1728--1735,
Sept. 1993.

\bibitem{G_20}
L. Zheng and D. N. C. Tse, ``Packing spheres in the Grassmann manifold:
A geometric approach to the noncoherent multiple-antenna channel''
IEEE Trans. Inform. Theory, vol. 48, pp. 359--383, Feb. 2002.

\bibitem{G_7}
A. Gersho, ``Asymptotically optimal block quantization'' IEEE Trans.
Inform. Theory, vol. IT-25, pp. 373--380, July 1979.

\bibitem{G_8}
H. Gish and J. N. Pierce, ``Asymptotically efficient quantizing'' IEEE
Trans. Inform. Theory, vol. IT-14, pp. 676--683, Sept. 1968.

\bibitem{G_14}
A. Kuh and B.W. Dickinson, ``Information capacity of associative memories''
IEEE Trans. Inform. Theory, vol. 35, pp. 59--68, Jan. 1989.

\bibitem{G_3}
J. Beirlant, E. J. Dudewicz, L. Gyorfi, and E. van der Meulen, ``Nonparametric
entropy estimation: An overview'' Int. J. Math. Stat. Sci.,
vol. 6, no. 1, pp. 17--39, June 1997.

\bibitem{G_1}
I. A. Ahmad and P.-E. Lin, ``A nonparametric estimation of the entropy
for absolutely continuous distributions,'' IEEE Trans. Inform. Theory,
vol. IT-22, pp. 688--692, May 1976

\bibitem{G_17}
A. Mokkadem, ``Estimation of the entropy and information of absolutely
continuous random variables,'' IEEE Trans. Inform. Theory, vol. 35, pp.
193--196, Jan. 1989.

\bibitem{Shore}
J. E. Shore, R. W. Johnson
``Axiomatic Derivation of the Principle of Maximum Entropy and the Principle of Minimum Cross-Entropy''
IEEE Trans. Inform. Theory, vol. 26, no. 1, pp. 26--37, Jan. 1980.

\bibitem{Amari}
S. Amari and H. Nagaoka, ``Methods of Information Geometry'' American Mathematical Society, Oxford University Press,
2000

\bibitem{PhysScripta}
S. A. Ali, C. Cafaro, A. Giffin, D. -H. Kim,
``Complexity Characterization in a Probabilistic Approach to Dynamical Systems Through Information Geometry and Inductive Inference'' Physica Scripta, 85(2), Febr. 2012.

\bibitem{Lancosh_8}
Lanczos K., ``The Variational Principles of Mechanics'' Toronto, 1962.

\bibitem{Plank_9}
Planck M., ``The unity of physical world views'' Physikalische Zeitschrift, Vol.10, pp. 62--75, 1909.

\bibitem{c65}
Jaynes E.T., ``Delaware Seminar in the Foundations of Physics'' ed. by M. Bun-ge. Berlin: Springer-Verlag, 1967.

\bibitem{c156}
Zupanovic P., Juretic D., Botric S.,
``Kirchhoff's loop law and the maximum entropy production principle''
Phys. Rev. E, V.70, pp. 056108, 2004.

\end{thebibliography}
\end{document}